\documentclass[aps,preprint]{revtex4-1}

\usepackage{graphicx}% Include figure files
\usepackage{dcolumn}% Align table columns on decimal point
\usepackage{bm}% bold math
\usepackage[dvipdfm]{hyperref}
\usepackage{amsmath}
\usepackage{amsfonts}
\begin{document}

\title{Three-dimensional Models of Topological Insulator Films: 
Dirac Cone Engineering and Spin Texture Robustness}

\date{\today}
\author{David Soriano$^{1*}$}
\author{Frank Ortmann$^1$}\thanks{These authors contributed equally.}
\author{Stephan Roche$^{1,2}$}
\affiliation{$^1$CIN2 (ICN-CSIC) and Universitat Aut\`onoma de Barcelona, Catalan Institute of Nanotechnology, Campus de la UAB, 08193 Bellaterra (Barcelona), Spain \\
$^2$ICREA, Instituci\'o Catalana de Recerca i Estudis Avan\c{c}ats, 08010 Barcelona, Spain}

\begin{abstract}
We have designed three-dimensional models of topological insulator thin films, showing a tunability of the odd number of Dirac cones on opposite surfaces driven by the atomic-scale geometry at the boundaries. This enables creation of a single Dirac cone at the $\Gamma$-point as well as possible suppression of quantum tunneling between Dirac states at opposite surfaces (and gap formation), when opposite surfaces are geometrically differentiated.  The spin texture of surface states was found to change from a spin-momentum-locking symmetry to a progressive loss in surface spin polarization upon the introduction of bulk disorder, related to the penetration of boundary states inside the bulk. These findings illustrate the richness of the Dirac physics emerging in thin films of topological insulators and may prove utile for engineering Dirac cones and for quantifying bulk disorder in materials with ultraclean surfaces.              
\end{abstract}


\maketitle

\emph{Introduction.-} The nascent field of Topological Insulators (TI) sparked by the seminal paper of Kane and Mele \cite{Kane:2005}, together with the prediction of three-dimensional structures for TI \cite{Bernevig:2006a}, and the subsequent experimental discoveries of two-dimensional HgCdTe quantum wells\cite{Konig:2007} and three-dimensional TI (3D-TI) materials \cite{Hsieh:2008, Hsieh:2009, Xia:2009, Zhang:2009,Yazyev:2010} has thrusted these fascinating materials to the forefront of modern condensed matter physics \cite{Hasan:2010, Hasan:2011, Qi:2011}. Topological insulators are governed by strong spin-orbit coupling and special crystalline symmetries that yield an insulating bulk phase complemented by highly robust, gapless Dirac boundary states, revealed through spin-resolved ARPES profiles, or through peculiar Landau levels fingerprints in scanning tunneling spectroscopy (STM) measurements \cite{Cheng:2010, Hanaguri:2010,Liu:2012}. However, despite the success in identifying these chiral surface states by photoemission and STM, the nature of surface transport in 3D TI lacks experimental characterization. This is because in all the materials studied to date, residual conduction through bulk states is irremediably driven by unintentional doping introduced by the electrical gates and contacts\cite{Kim:2011,Steinberg:2011}.

The aforementioned boundary states are described by Dirac-cone physics, similarly to the case of low-energy excitations in graphene \cite{CastroNeto:2009}, but with Dirac cones appearing in odd numbers. The robustness of the physics of these chiral states, with respect to the thickness of a TI film, deserves particular attention.  Indeed when TI are reduced to thin films, quantum tunneling between Dirac states at opposite surfaces can eventually occur, yielding gap formation, as recently shown for  $ {\rm Bi}_{2}{\rm Se}_{3}$ \cite{Taskin:2012} or freestanding thin Sb films \cite{Bian:2012}. However, and surprisingly, specific interactions between film and substrate prevent gap formation \cite{Bian:2012},  a feature which could be of considerable interest for spintronic applications, but which remains poorly understood.

Understanding the effects of disorder on quantum transport of massless Dirac fermions is a challenging but fundamental task. For a single scattering event, the spin (or pseudospin, for graphene) quantum degree of freedom may lead to partial or full suppression of backward reflection when the charge crosses a local tunneling barrier (referred to as the Klein tunneling mechanism \cite{Ando:1998, Katsnelson:2006, Moore:2010}). Additionally, quantum interferences between propagating trajectories may lead to an increase in the semiclassical conductivity monitored by the $\pi$ Berry phase (weak antilocalization) \cite{McCann:2006,Shen:2011a,Shen:2011b}. These mechanisms prevent the transition to a strong Anderson localization regime and vanishing conductivity; their dependence on the nature and strength of disorder demands in-depth scrutiny. 

All types of non-magnetic disorders, including structural imperfections (e.g. vacancies), surface contaminants, or doping with chemical impurities\cite{Noh:2011, Shan:2011, Schaffer:2012, Schubert:2012},  preserve time-reversal symmetry and are expected to weakly affect TI transport physics. In contrast, magnetic impurities (which break time reversal invariance) can develop net magnetic moments inducing local magnetic ordering, spin-dependent scattering or gap formation \cite{Culcer:2012,Caprara:2012,Rosenberg:2012, Foster:2012,Honolka:2012,Ye:2012}. Henk and coworkers \cite{Henk:2012} recently reported on the robustness of Dirac states upon moderate Mn doping of a $ {\rm Bi}_{2}{\rm Te}_{3}$ surface layer. However, they observed complicated spin textures for both undoped as well as Mn-doped $ {\rm Bi}_{2}{\rm Te}_{3}$, which exhibited layer-dependent spin reversal and spin vortices.  Since the topological protection of Dirac states is inherently driven by the non-trivial topology of bulk electronic wavefunctions, surface and bulk disorders are actually expected to tailor spin polarization features. However, these effects and their relation to the eventual Anderson localization of Dirac fermions have yet to be quantified. 
 
In this Letter, we describe 3D models of TI thin films and show that Dirac-cone characteristics on opposite surfaces of the film can be tuned upon differentiation of atomic-scale surface terminations.  Reducing the film thickness to several bulk layers leads to a loss of low-energy Dirac physics, owing to quantum tunneling between chiral states lying at opposite surfaces. In striking contrast, when atomic-scale bottom and top surfaces are geometrically differentiated, Dirac cones develop either at the $\Gamma$-point (single Dirac cone) or at M-points (triple Dirac cones) and remain uncoupled down to a few bulk-layers. Our findings are consistent with recent experimental observations \cite{Bian:2012} and open the way to controlled engineering of thin 3D-TI with highly robust chiral states. Furthermore, upon analyzing the spin textures of Dirac states on surfaces of thick TI films as a function of the strength of non-magnetic bulk disorder, we found that disorder leads to steady randomization of polarization properties and to suppression of certain spin-momentum locking symmetries.

\vspace{1em}

\begin{figure}[t]
\includegraphics[width=\linewidth,angle=0]{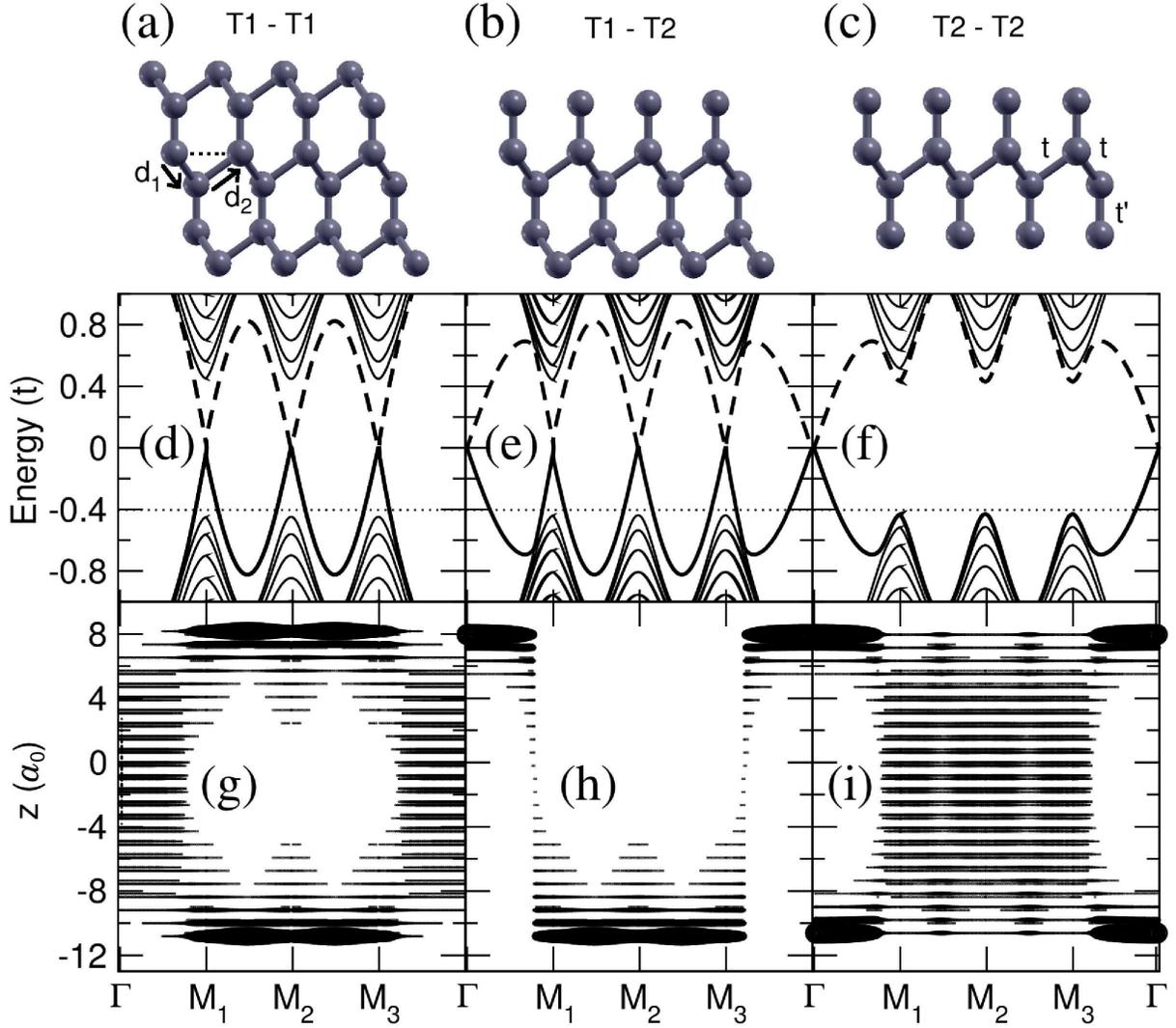}
\caption{ \label{fig1}(a-c) Thin slabs showing the atomic structure of the top and bottom surfaces for the T1--T1, T1--T2 and T2--T2 slab geometries. (d-f) Band structure along the path $\Gamma$ - M$_1$ - M$_2$ - M$_3$ - $\Gamma$ for the three slab geometries shown in (a-c). In the T1--T2 case the bands are no longer degenerated due to inversion symmetry breaking. (g-i) Charge density plots corresponding to the degenerate valence bands $|\Psi_{VB}(\mathbf{k})|^2$  and $|\Psi'_{VB}(\mathbf{k})|^2$ obtained for each slab geometry.    
   }
\end{figure}

\emph{Model.-} To describe the 3D-TI films, we used the Fu--Kane--Mele (FKM) Hamiltonian which is defined on a diamond lattice with a single orbital per site. \cite{Fu:2007} This is a three-dimensional generalization of the model proposed by Kane and Mele to study the quantum spin Hall (QSH) effect in two-dimensional honeycomb lattices in the presence of spin-orbit coupling (SOC) \cite{Kane:2005,Kane2:2005}   
\begin{equation}
\mathcal{H} = t\sum_{\langle ij \rangle} c_i^\dagger c_j + i(8\lambda_{SO}/a^2)
\sum_{\langle\langle ij \rangle\rangle} c_i^\dagger \mathbf{s}\cdot(\mathbf{d}_{ij}^1
\times \mathbf{d}_{ij}^2) c_j.
\label{eqn1}
\end{equation}
The first term denotes the hopping term ($t>0$) between nearest neighboring orbitals, while the second describes the spin-orbit interaction (SOI) given by a spin-dependent complex term connecting second neighbors $i$ and $j$ in the diamond structure through vectors $\mathbf{d}^1_{ij}$ and $\mathbf{d}^2_{ij}$ along first-neighbor bonds (see Fig.\ref{fig1}(a)). $\lambda_{SO}$ is the SOI strength, $a$ is the cubic cell size and $\mathbf{s} = (\sigma^x,\sigma^y,\sigma^z)$ is elaborated from the Pauli matrices. From the diamond bulk Hamiltonian we create slabs with varying number of layers (up to 48) and (111) surface orientation. 

An important feature of the FKM model is that it enables the description of either a weak or a strong topological insulator depending on the value of the hopping $t'$ along the (111) direction (see Fig.\ref{fig1}(c)). When $t'<t$, a weak topological insulating phase is generated whose physics resembles that of stacked bilayer bismuth, where each layer is in a 2D QSH state \cite{Murakami:2006}. This phase is characterized by an even number of Dirac points in the surface Brillouin zone (SBZ). Alternatively, if $t'>t$, the system is driven into a strong topological insulating (STI) phase, with an odd number of Dirac cones centered at the M-points in the SBZ. Such a phase has been found for instance in Bi$_{1-x}$Sb$_x$ \cite{Hsieh:2009}.

Here we focus on the STI phase, which seems more relevant in light of recent experimental works \cite{Hsieh:2008,Hsieh:2009,Chen:2009,Xia:2009,Xu:2011,Beidenkopf:2011,Valla:2012,Taskin:2012}, and show that a gap can form at the M points (by removing the outermost layers of a diamond slab) and create a new surface Dirac cone at the $\Gamma$- point \cite{Hasan:2010} showing a band inversion at $\mathbf{k} = 0$. This extends the applicability of the FKM model to strong topological insulators such as Bi$_2$Se$_3$, Bi$_2$Te$_3$ or Sb$_2$Te$_3$ and could enable further modeling of some ternary \cite{Zhang:2011,Dai:2012} and non-centrosymmetric \cite{Bahramy:2012} compounds.

\vspace{1em}

\emph{Dirac cone engineering by atomic-scale surface geometry differentiation.-} Fig.\ref{fig1}(d-f) show the band structures of the three different diamond films obtained by varying surface terminations. We have  fixed $t=1$, $t'=1.4$, $\lambda_{SO}=1/8$, and have used a unit cell with $N=48$ sites (one site per layer) which corresponds to a slab thickness of about $19 a_0$ (where $a_0 = a/\sqrt{2}$ is the distance between second neighbor atoms in the diamond lattice). The slab with standard termination in both surfaces (see Fig.\ref{fig1}(a)) has already been investigated by Fu and coworkers in Ref.\cite{Fu:2007}. We label this termination T1, and the related slab geometry, T1--T1. The band structure for this T1--T1 geometry shows three Dirac cones located at the three equivalent M points in the SBZ (Fig.\ref{fig1}(d)). Since this geometry preserves inversion symmetry, all the bands are two-fold degenerated with three Dirac cones located on each surface as shown by the charge density plot corresponding to the degenerate valence bands  $|\Psi_{VB}(\mathbf{k})|^2$  and $|\Psi'_{VB}(\mathbf{k})|^2$ (Fig.\ref{fig1}(g)), calculated along the path $\Gamma$--M$_1$--M$_2$--M$_3$--$\Gamma$.   

Removing the uppermost layer from the top surface (while keeping the second-neighbor hopping within the layer underneath) one generates another termination labeled T2 (see Fig.\ref{fig1}(b)). The corresponding slab geometry with differing terminations is labeled T1--T2 and its band structure exhibits four Dirac cones along the path $\Gamma$--M$_1$--M$_2$--M$_3$--$\Gamma$ (Fig.\ref{fig1}(e)). Although the number of Dirac cones is even the system remains in the STI phase because the number of Dirac points on each surface remains odd. In fact, the three Dirac cones located at the M points are related to the surface with T1 termination, while the single Dirac cone (emerging at the $\Gamma$ point) is localized at the T2 terminated surface, as evidenced by the valence band charge density plot (Fig.\ref{fig1}(h)). In such a T1--T2 slab geometry, the inversion symmetry is broken and the valence and conduction bands are no longer degenerated. However, the topological states in both surfaces can be continuously transformed from one to another by tuning the wave vector. This fact has important implications regarding film thickness, as we explain below. 

A third possible structure is obtained by also removing one layer from the bottom surface, which leads to a T2--T2 slab geometry (Fig.\ref{fig1}(c)). The electronic structure of this slab includes a single Dirac cone at the $\Gamma$ point in the SBZ of each surface (Fig.\ref{fig1}(f)) and resembles the typical band structure of topological insulators like Bi$_2$Se$_2$ \cite{Xia:2009} or Bi$_2$Te$_3$.\cite{Chen:2009} The valence band charge density plot shows that the states at the Dirac cones are localized at opposite surfaces (Fig.\ref{fig1}(i)).     

\begin{figure}[t]
\includegraphics[width=0.9\linewidth,angle=0]{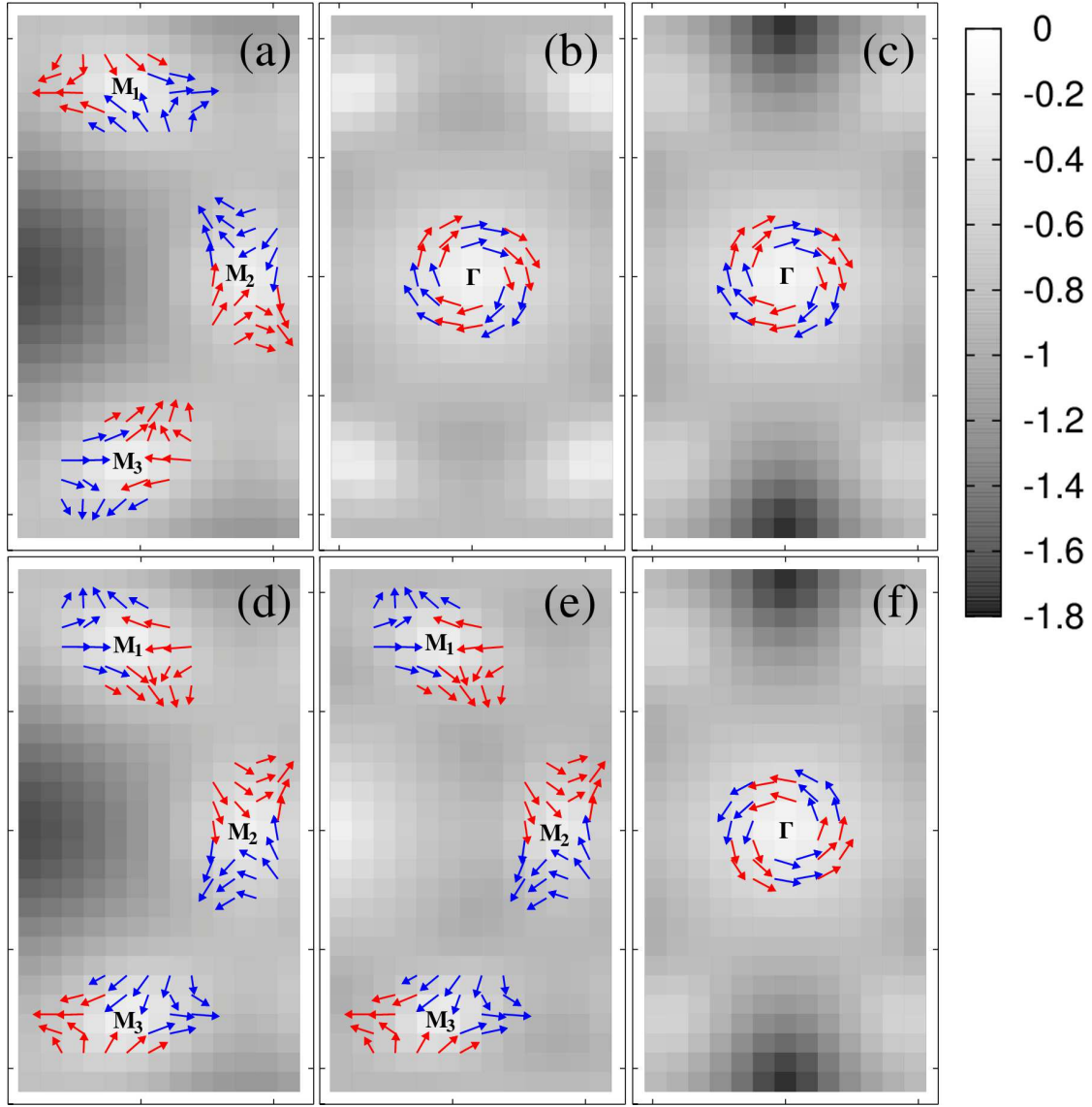}
\caption{ \label{fig2} (Color). Spin texture on top and bottom surfaces (top and bottom pannels respectively) for the three different slabs studied (a,d) T1--T1, (b,e) T1--T2 and (c,f) T2--T2. The gray scale indicates the valence band energy where the Dirac cones are located at the bright spots.   
   }
\end{figure}

We now compute the spin texture on each surface for the three slab geometries by evaluating the expectation value of the spin operator $\langle \mathbf{\hat{s}} \rangle$ of the corresponding surface valence band state $\Psi_{VB}(\mathbf{k})$ projected onto the surface sites $i$ 
\begin{equation}
\mathbf{S}(\mathbf{k}) = \sum_{\substack{ 
   i \in surf \\
   \tau,\tau '
  }} \langle \Psi_{VB}(\mathbf{k}) | i,\tau \rangle \langle i,\tau | \mathbf{\hat{s}} | i,\tau ' \rangle \langle i,\tau ' | \Psi_{VB}(\mathbf{k}) \rangle
\label{eqn2}
\end{equation}
where $\tau,\tau '$ are spin indices.  

In Fig.\ref{fig2}, we superimpose the spin textures (restricted to $E/t=-0.4 \pm 0.1$, see dotted line on Fig.\ref{fig1}(d-f)) on the valence band energy where Dirac points correspond to the brightest areas ($E=0$). The top and bottom figures correspond to the top and bottom surfaces, respectively. Blue and red arrows respectively correspond to negative and positive $z$-components (out-of-plane) of the spin. For the three studied surface terminations, the states around the Dirac points show an out-of-plane helical spin texture preserving time-reversal symmetry. It is noteworthy that in the T1--T1 and T2--T2 cases, the spin polarizations in the two surfaces are related by inversion symmetry and exhibit a vortex or a spin reversal texture. This is not the case for the T1--T2 case, in which the inversion symmetry is broken and the spin texture is centered around the $\Gamma$ point (at the top surface) and the three M points (at the bottom surface).      

\begin{figure}[t]
\includegraphics[width=\linewidth,angle=0]{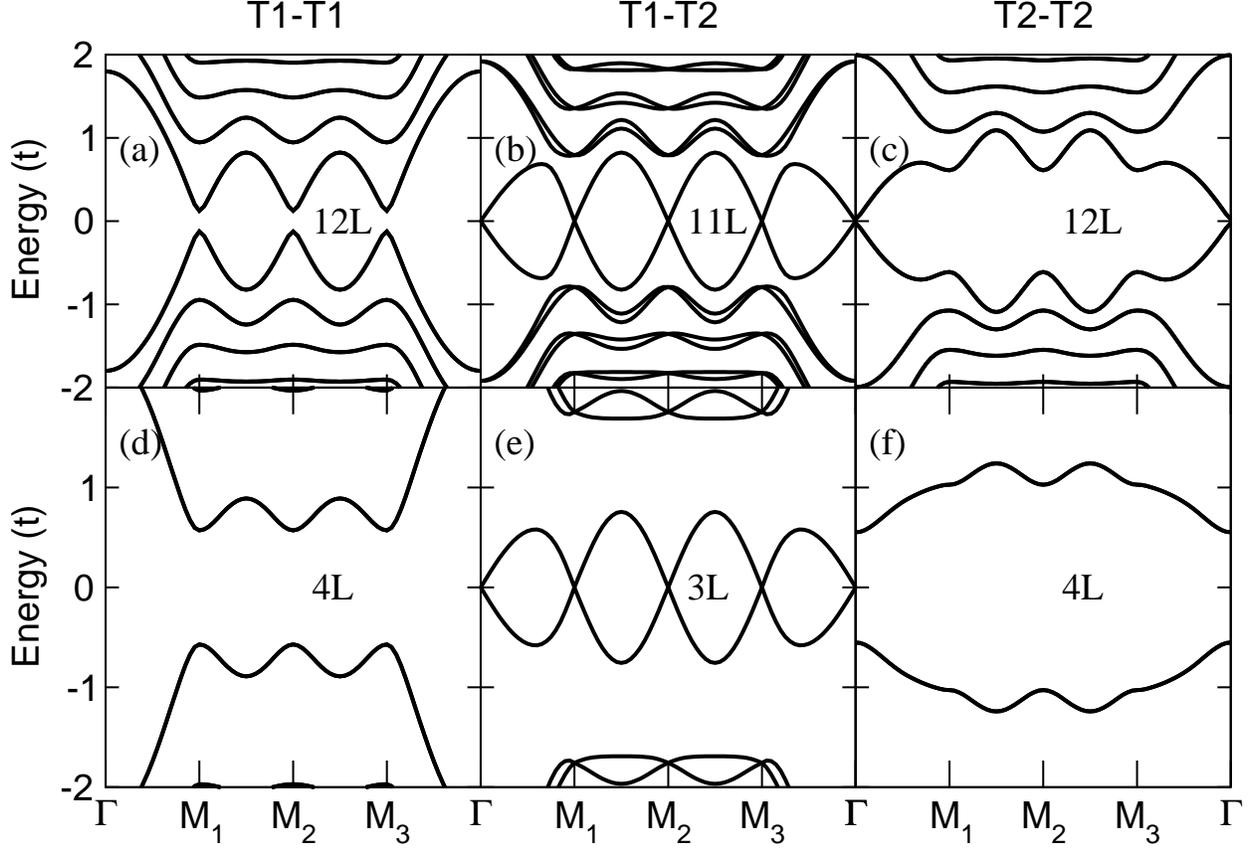}
\caption{ \label{fig3}  Band structure of slabs of various thicknesses (layers L) and surface terminations made from the STI phase as explained in the text. The surface terminations of upper and lower surfaces are T1--T1 (a,d), T1--T2 (b,e), and T2--T2 (c,f).
  }
\end{figure}

\vspace{1em}

\emph{TI-film thickness and robustness of Dirac physics.-} Fig.\ref{fig3}  gives the electronic structures of slabs with varying film thickness. For the T1--T1 terminated structure, a sizable gap already opens at all M points for slabs with twelve layers. However, reducing the thickness down to four layers (Fig.\ref{fig3} (d)) provides  insulating surface states, since gap values are above one (in $t$ unit). For the T2--T2 case, the twelve-layer slab evidences a small gap at $\Gamma$ which is further widened upon reduction of film thickness (Fig.\ref{fig3} (c,f)). Note that a similar situation has been encountered in recent experiments. \cite{Zhang:2010,Taskin:2012} For four layers, a gap of approximately one (in $t$ unit) develops, similarly to the T1--T1 termination. Turning to the mixed (T1--T2) termination, one might expect a similar trend. However, the behavior is completely different (see Fig.\ref{fig3} (b,e)): gapless surface states are insensitive to quantum tunneling and gap formation is suppressed regardless of film thickness. These results support the interpretation of recent experiments by Bian and coworkers \cite{Bian:2012},  who also reported the absence of a gap opening for thin TI-films. In that case strong interfacial bonding to the substrate prevents gap opening, in contrast to freestanding TI-films.

\begin{figure}[t]
\includegraphics[width=\linewidth,angle=0]{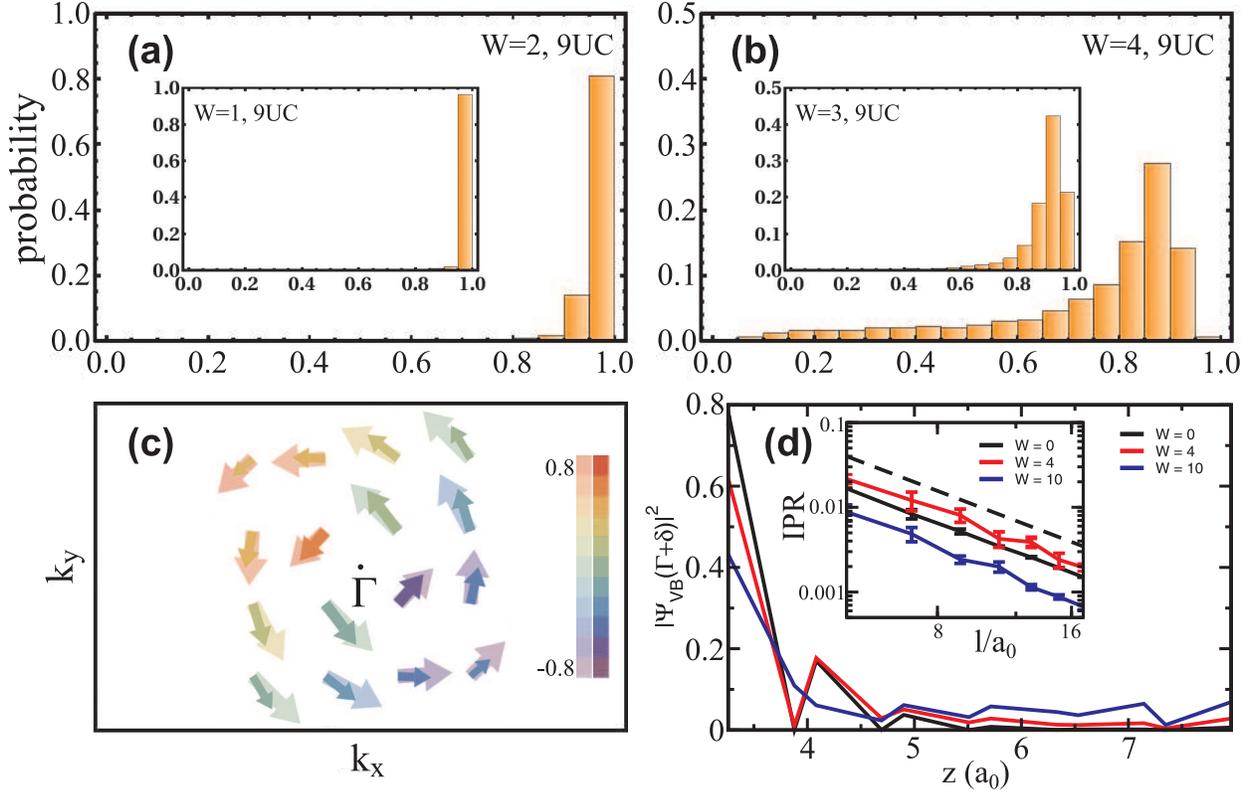}
\caption{(Color). Spin polarization of T2--T2 slab (12 layers) with disorder. (a,b) norm distribution of spin vectors close to $\Gamma$. (c) Spin texture of clean (semi-transparent) and disordered slab (opaque). (d) Squared valence band wave-function near the $\Gamma$-point for different disorder and inverse participation ratio (IPR) for increasing lateral slab size (inset). The $l^{-2}$ scaling behavior is shown in dashed lines for comparison.}
\label{fig4}
\end{figure}

\vspace{1em}

\emph{Bulk disorder effects on spin textures.-} We investigate the changes in the spin texture upon introduction of bulk disorder. It is indeed very instructive to determine the extent to which the topological protection of surface states is reduced in the presence of bulk disorder with increasing strength. To facilitate comparison with experiments, we focus on the geometry which induces single $\Gamma$-centered Dirac points (T2--T2 surface configuration). For each k-point close to $\Gamma$, the spin vectors $\mathbf{S}(\mathbf{k})$ are computed (using Eq.\ref{eqn2} but neglecting surface projection) for the valence band. The spin textures are plotted for the clean case in Fig.\ref{fig4}(c) (semi-transparent arrows) where the in-plane projection is indicated by the vector length and the out-of plane component by the color index.  For all k-points in the vicinity of $\Gamma$, the total length is one (in units of $\hbar/2$). 

The introduction of bulk disorder (excluded from surface layers) is found to alter surface spin textures. To monitor these changes, we compute the total norm of the spin vectors taken from a regular grid of k points and a series of (up to 100) different disorder configurations, and then plot their statistics in histograms (Fig.\ref{fig4}). A generic model of Anderson disorder \cite{Anderson:1958} is included through a modulated potential profile ($\varepsilon_{i}$), with $\varepsilon_{i}$ selected at random in the interval $[-W/2,W/2]$. This model mimics impurities or structural defects and has been widely employed in the literature for exploring metal-insulator transition \cite{Schubert:2012,Lee:1981,Evers:2008}. We use a $12$ layers thick supercell containing $9$ (i.e. $3 \times 3$) unit cells (UC).  Complementary data for different supercells are discussed in the supplementary material\cite{supp}. 

Upon varying the disorder strength $W$ (Fig.\ref{fig4}(a,b)), we find that as long as $W\leq 1$, the spin textures remain unchanged by bulk disorder, whereas starting from $W\simeq 2$, spin polarization starts being randomized with a reduced norm (for $W\geq 4$) which eventually vanishes in the strong disorder limit. An illustration of the randomization and loss of spin-polarization for $W=4$ and 9UC is given in Fig.\ref{fig4}(c) (opaque arrows) in comparison to the clean case. 

To unveil the mechanisms leading to the randomization of the spin-polarization, we plot the absolute square of the valence band wavefunction $|\Psi_{VB}(\mathbf{k})|^2$ near the $\Gamma$-point ($\mathbf{k} = \Gamma + \delta$) along the $z$-axis perpendicular to the surface (Fig.\ref{fig4}(d)). For $W = 0$ the valence band wavefunction is mainly localized at the surface, but as the disorder strength is tuned from $W = 4$ to $W=10$, it further spreads over bulk layers. The observed penetration depth of electronic states progressively increases with $W$,  changing the nature of wavefunctions from (quasi) two-dimensional confined states at the surface to more three-dimensional real-space extended states promoted by bulk disorder. For W=10, the wavefunction is seen to be spread all over the system. However we also observe that inter-surface coupling mediated by bulk disorder is a minor effect. In fact, from a comparison of spin polarization histograms for different slab thicknesses (12-layer, 22-layer and 46-layer slabs)\cite{supp} we conclude that inter-surface coupling between Dirac cones is negligible for spin randomization at thicknesses down to 12 layers.

To deepen the analysis, we compute and analyze the scaling behavior of the inverse participation ratio defined as ${\rm IPR} =\sum_i |\Psi_{VB}|^4/([\sum_i |\Psi_{VB}|^2]^2)$. The IPR is a common measure for the localization nature of electronic states (see e.g. Brndiar \emph{et al} \cite{Brndiar:2008}).  In absence of disorder, the IPR is predicted to scale as $l^{-d}$ (where $l$ is system length, $d$ the space dimension) being a fingerprint of truly extended states and a metallic regime, whereas the Anderson localization regime (in the strong disorder limit) manifests in a length-independent IPR value (with ${\rm IPR} \sim \xi^{-d}$, $\xi$ the localization length).  Fig.\ref{fig4}-d (inset) shows ${\rm IPR}$ for increasing lateral slab sizes $l$ and for $W=0$, $4$ and $10$.  It is found that ${\rm IPR}\sim l^{-2}$ for $W=0$, in agreement with extended wavefunctions at the surface (only weak unavoidable spreading to nearest bulk layers is observed for $|\Psi_{VB}|^2$, Fig.\ref{fig4}-d (main frame)). By increasing the bulk disorder strength from $W=4$ to $W=10$, the IPR are seen to vary in absolute value in a non-monotonic fashion while maintaining the IPR$(l)\sim l^{-2}$ scaling behavior, with no sign of saturation for the considered system sizes. This scaling analysis excludes short localization lengths and Anderson insulating regime for bulk disorder strengths which however significantly suppress surface spin polarization. 

\vspace{1em}

\emph{Conclusion.-} We have generated models of 3D TI films with varying thickness, surface termination and tailored Dirac-cone characteristics. Anomalously robust Dirac cones in ultrathin TI have been obtained when opposite surfaces are structurally differentiated, a feature that has also been recently reported experimentally although attributed to chemical differentiated surfaces \cite{Bian:2012}. Additionally, a scaling analysis of bulk-disorder effects on spin polarization (in a thick-film model exhibiting single surface Dirac cones at the $\Gamma$-point) has revealed that spin randomization is concomitant to a penetration of boundary states into the bulk. These findings (of relevance for real materials such as Bi$_2$Se$_3$ and related TI) suggest ways to analyze the bulk crystalline quality of TI by inspecting the spin texture fingerprints (through spin-resolved ARPES) at ultraclean TI-surfaces, in a regime far form the Anderson insulating regime.

%\bibliography{gamma-FKM_SUB.bib}

%

\end{document}